\documentstyle[preprint,aps,psfig]{revtex}
\begin{document}
\draft
\preprint{MKPH-T-97-1}
\title{{ \bf On the Gerasimov-Drell-Hearn sum rule for the deuteron}}

\author{H.\ Arenh\"ovel, G.\ Kre\ss, R.\ Schmidt and P.\ Wilhelm}
\address{
Institut f\"{u}r Kernphysik, Johannes Gutenberg-Universit\"{a}t,
       D-55099 Mainz, Germany}
\maketitle

\begin{abstract}
The Gerasimov-Drell-Hearn sum rule is evaluated for the deuteron by 
explicit integration up to a photon 
energy of 550 MeV including contributions from the photodisintegration 
channel and from coherent and incoherent single pion production as well. The 
photodisintegration channel converges fast enough in this energy range and 
gives a large negative contribution, essentially from the $^1S_0$ resonant 
state near threshold. Its absolute value is about the same size than the 
sum of proton and neutron GDH values. It is only partially cancelled by the 
single pion production contribution. But the incoherent channel 
has not reached convergence at 550 MeV. 
\end{abstract}

\pacs{PACS numbers: 11.55.Hx, 24.70.+s, 25.20.Dc, 25.20.Lj}

\section{Introduction}
\label{sec1}

The Gerasimov-Drell-Hearn (GDH) sum rule connects the anomalous magnetic moment 
of a particle with the energy weighted integral - henceforth denoted by 
$I^{GDH}$ - from threshold up to infinity over the difference of the total 
photoabsorption cross sections for circularly polarized photons on a target 
with spin parallel and antiparallel to the spin of the photon. 
In detail it reads for a particle of mass $M$, charge $eQ$, anomalous 
magnetic moment $\kappa$ and spin $S$ 
\begin{equation}
I^{GDH}=4\pi^2\kappa^2\frac{e^2}{M^2}\,S
=\int_0^\infty \frac{dk}{k}
\left(\sigma ^P(k)-\sigma ^A(k)\right)
\,,\label{gdh}
\end{equation}
where $\sigma ^{P/A}(k)$ denote the total absorption cross sections for
circularly polarized photons on a target with spin parallel and antiparallel 
to the photon spin, respectively, and the anomalous magnetic moment is 
defined by the total magnetic moment operator of the particle 
\begin{equation}
\vec M = (Q+\kappa)\frac{e}{M}\vec S\,.
\end{equation}

This sum rule gives a very interesting relation between a 
magnetic ground state property of a particle and an integral property of 
its whole excitation spectrum. In other words, this sum rule shows that the 
existence of a nonvanishing anomalous magnetic moment points directly to an 
internal dynamic structure of the particle. Furthermore, because the lhs 
of (\ref{gdh}) is positive, it tells us that the integrated, energy-weighted 
total absorption of a circularly polarized photon on a particle with its 
spin parallel to the photon spin is bigger than the one on a target with 
its spin antiparallel, if the particle posesses a nonvanishing anomalous 
magnetic moment.

The GDH sum rule has first been derived by Gerasimov \cite{Ger65} and, 
shortly afterwards, independently by Drell and Hearn \cite{DrH66}. It is 
based on two ingredients which follow from the general principles of 
Lorentz and gauge invariance, unitarity, crossing symmetry and causality 
of the Compton scattering amplitude of a particle. The first one is the 
low energy theorem of Low \cite{Low54} and Gell-Mann and Goldberger 
\cite{GeG54} for a spin one-half particle which later has been generalized 
to arbitrary spin \cite{LaC60,Sai69,Fri77}. 
The second ingredient is the assumption of an unsubtracted dispersion 
relation for the difference of the elastic forward scattering
amplitudes for circularly polarized photons and a completely polarized
target with spin parallel and antiparallel to the photon spin. 

Since proton and neutron have large anomalous magnetic moments, one finds 
large GDH sume rule predictions for them, i.e., $I^{GDH}_p=204.8\,\mu$b 
for the proton and $I^{GDH}_n=233.2\,\mu$b for the neutron. 
Although this sum rule is known for more than 30 years, it has never been 
verified by a direct measurement. Early evaluation by Karliner \cite{Kar73}, 
based on a multipole analysis of experimental data on pion 
photoproduction on the nucleon, did not give 
conclusive results due to the lack of data at higher energies, and even 
present day data do not allow a definite answer as to its validity (see e.g. 
\cite{SaW94}). The recent interest in this sum rule stems from the 
study of the spin dependent structure functions in deep inelastic 
scattering \cite{Dre95}. 

Applying the GDH sume rule to the deuteron, one finds a very interesting 
feature. On the one hand, the deuteron has isospin zero ruling out the 
contribution of the large nucleon isovector anomalous magnetic moments to 
its magnetic moment. Therefore, one expects a very small 
anomalous magnetic moment for the deuteron. In fact, the experimental value 
is $\kappa_d=-.143$ resulting in a GDH prediction of $I^{GDH}_d = 0.65\,\mu$b, 
which is more than two orders of magnitude smaller 
than the nucleon values. On the other hand, considering the possible 
absorption processes, we note first that the incoherent pion production on 
the deuteron is dominated by the quasifree production on the nucleons bound 
in the deuteron. Thus it is plausible to expect from these processes a 
contribution to the GDH value roughly given by the sum of the proton and 
neutron GDH values, i.e., 438 $\mu$b. Additional contributions arise from 
the coherent $\pi^0$ production channel. In order to obtain the small total 
deuteron GDH value one, therefore, needs a large negative contribution of 
about the same size for cancellation. Indeed, one has an additional channel not 
present for the nucleon, namely the photodisintegration channel which is 
the only photoabsorption process below the pion production threshold. 
A closer look shows in fact that at very low energies near threshold a 
sizeable negative contribution arises from the $M1$-transition to the 
resonant $^1S_0$ state, because this state can only be reached if the spins 
of photon and deuteron are antiparallel, and is forbidden for the parallel 
situation. 

It is the aim of the present paper to report on an evaluation of the GDH sum 
rule for the deuteron by explicit integration of the GDH integral up to a 
photon energy of 550 MeV including the photodisintegration channel as well 
as coherent and incoherent single pion photoproduction channels.

\section{The GDH sum rule for the deuteron}
\label{sec3}
For the deuteron, one can express the difference of the cross sections by 
the vector target asymmetry $\tau^c_{10}$ \cite{ArS91}, i.e., 
\begin{equation}
\sigma ^P(k)-\sigma ^A(k)=\sqrt{6}\sigma^0 \tau^c_{10}\,,
\end{equation}
where $\sigma^0 $ denotes the unpolarized absorption cross section. 
Explicitly, $\sigma^0 $ and the difference may be expressed in terms of 
the electric and magnetic multipole matrix elements by
\begin{eqnarray}
 \sigma^0&=& \frac{4\pi}{9}\sum_{Lj \lambda }\Big( |E^L(\lambda j)|^2
             + |M^L (\lambda j)|^2\Big)\,,\label{sig0}\\
 \sigma ^P(k)-\sigma ^A(k)&=& 4\pi\sqrt{6} 
 \sum_{LL'j \lambda } (-)^{j}
 \left(\matrix{L'&L& 1 \cr 1 &-1&0 \cr}\right)
 \Bigg\{ \matrix{L'&L&1 \cr 1&1&j \cr} \Bigg\} \nonumber\\
 & &\qquad\qquad \Re e[(E^{L'} (\lambda j)+M^{L'} (\lambda j))^*
 (E^L (\lambda j)+M^L (\lambda j))]\,,\label{multd}
\end{eqnarray}
where $\lambda$ labels the possible final states of a given total angular 
momentum $j$. Note, that due to parity conservation one has in (\ref{sig0}) 
and (\ref{multd}) either electric or magnetic contributions for a given 
multipolarity $L$ and state $\lambda j$. 

We have evaluated explicitly the GDH sum rule for the deuteron by 
integrating the difference of the two total photoabsorption cross sections 
with photon and deuteron spins parallel and antiparallel up to a photon energy 
of 550 MeV. Three contributions have been included: (i) the 
photodisintegration channel $\gamma d \rightarrow n p$, (ii) the coherent 
pion production $\gamma d \rightarrow \pi^0 d$, and (iii) the 
incoherent pion production $\gamma d \rightarrow \pi N N$. The upper 
integration limit of 550 MeV has been chosen because on the one hand one finds 
sufficient convergence for the photodisintegration channel, while on the other 
hand, only single pion photoproduction has been considered, thus limiting the 
applicability of the present theoretical treatment to energies not too far 
above the two pion production threshold as long as significant 
contributions from multipion production cannot be expected. Indeed, the 
evaluation of $I^{GDH}$ for the nucleon by Karliner \cite{Kar73} indicates 
that significant contributions from two-pion production start only above 
this energy. We will now discuss the three contributions separately. 

\subsection{Photodisintegration}
\label{sec3a}
In this case, the final states are the partial waves of $np$ scattering, and 
for a fixed $j$, one has four final partial waves 
which are labelled by $\lambda$ in the Blatt-Biedenharn convention. For 
the leading contributions $L=L'=1$, one finds
\begin{eqnarray}
 \sigma ^P(k)-\sigma ^A(k)= -\frac{2\pi}{3}
\sum_{\lambda }&\Big(&2|E^1 (\lambda 0)|^2 +|E^1 (\lambda 1)|^2
                    -|E^1 (\lambda 2)|^2 \nonumber\\
&+&2|M^1 (\lambda 0)|^2 + |M^1 (\lambda 1)|^2  
     -|M^1 (\lambda 2)|^2\Big)\,.
\end{eqnarray} 
We first note, that $E1$-transitions lead to the $^1P_1$ and 
$^3P_j$ ($j=0,1,2$) states. However, the isoscalar $^1P_1$ is 
largely suppressed while the triplet $^3P_j$ contributions almost 
cancel each other. The cancellation would be complete if spin-orbit and 
tensor forces could be neglected, because in this case the matrix elements 
are simply related by angular momentum recoupling coefficients. Thus, at 
low energies remain the $M1$-transitions, essentially to $^1S_0$ and 
$^3S_1$ states. Of these, the $^1S_0$ contributions is dominant because 
of the large isovector part of the $M1$-operator coming from the large 
isovector anomalous magnetic moment of the nucleon. It is particularly 
strong close to break-up threshold where the 
$^1S_0$ state is resonant. It can only be reached by the antiparallel spin 
combination resulting in a strong negative contribution to the GDH sum rule. 

The photodisintegration channel is evaluated within 
the nonrelativistic framework as is described in detail in
Ref.~\cite{ArS91} but with inclusion of the most important relativistic 
contributions. Explicitly, all electric and magnetic multipoles up to 
the order $L=4$ are considered which means 
inclusion of the final state interaction in all partial waves up to $j=5$.
For the calculation of the initial deuteron and the final n-p scattering 
wave functions we use the realistic Bonn potential (r-space version) 
\cite{MaH87}. 
In the current operator we distinguish the one-body currents with Siegert 
operators (N), explicit meson exchange contributions (MEC) beyond the
Siegert operators, essentially from $\pi$- and $\rho$-exchange, 
contributions from isobar configurations of the wave functions (IC), 
calculated in the impulse approximation \cite{WeA78}, and leading order 
relativistic contributions (RC). 

\begin{figure}
\centerline{\psfig{figure=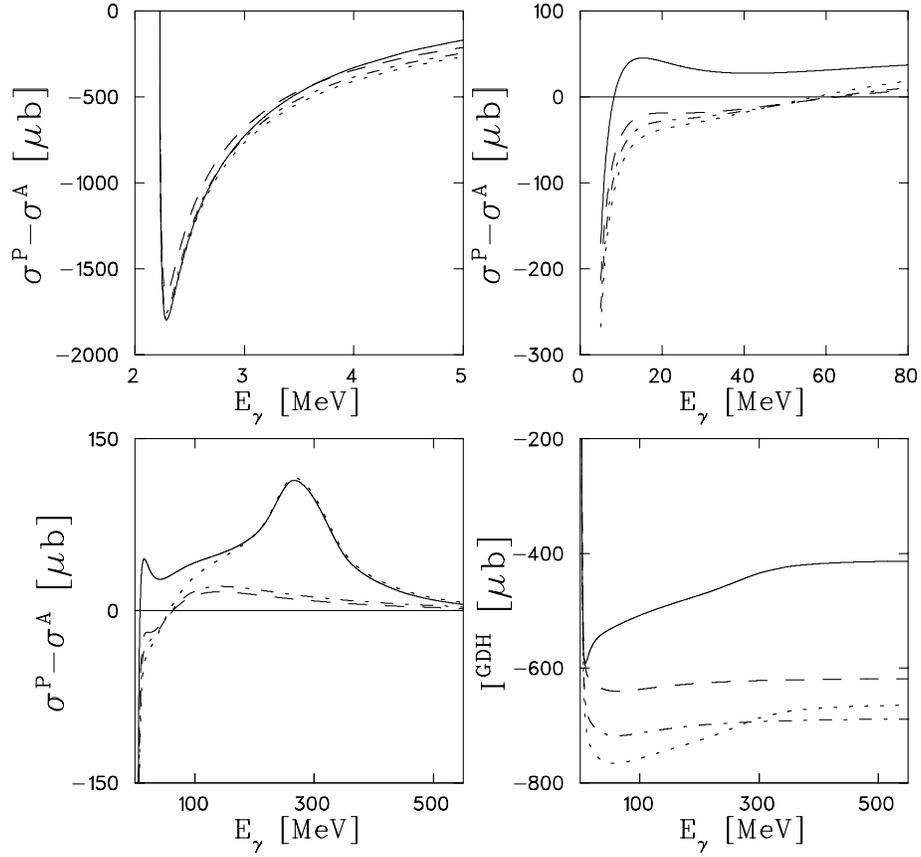,width=12cm,angle=0}}
\vspace*{.5cm}
\caption{
Contribution of the photodisintegration channel to the GDH sum rule for the 
deuteron. Two upper and lower left panels: difference of the cross sections 
in various energy regions; lower right panel: 
$I^{GDH}_{\gamma d \to np}$ as function of the upper integration energy. Dashed 
curves: N, dash-dot: N+MEC, dotted: N+MEC+IC, and full curves N+MEC+IC+RC.}
\label{fig1}
\end{figure}

The results are summarized in Fig.\ \ref{fig1}, where the cross section 
difference and the GDH integral is shown. The GDH values are listed in 
Tab.\ \ref{tabdis}. One readily notes the huge negative contribution from 
the $^1S_0$-state at low energies (see the upper left panel of Fig.\ 
\ref{fig1}). Here, the effects from MEC 
are relatively strong resulting in an enhancement of the negative value by 
about 15 percent. It corresponds to the well-known 10 percent enhancement 
of the radiative capture of thermal neutrons on protons. 
Isobar effects are significant in the region of 
the $\Delta$-resonance, as expected. They give a positive contribution, but 
considerably smaller in absolute size than MEC. The largest positive 
contribution stems from RC in the energy region up to about 100 MeV 
(see the upper right panel of Fig.\ \ref{fig1})
reducing the GDH value in absolute size by more than 30 percent. This 
strong influence from relativistic effects is not surprising in view of the 
fact, that the correct form of the term linear in the photon momentum of 
the low energy expansion of the forward Compton scattering amplitude 
is only obtained if leading order relativistic contributions 
are included\cite{Fri77}. 
The total sum rule value from the photodisintegration channel then is 
$I^{GDH}_{\gamma d \to np}(550\,\mbox{MeV})=-413\,\mu$b. Almost the same 
value is obtained for the Paris Potential \cite{Kre96}. Its absolute value 
almost equals within less than ten percent the sum of the free proton 
and neutron values. This may not be accidental since the large value is 
directly linked to the nucleon anomalous magnetic moment as is demonstrated 
by the fact that one finds indeed a very small but positive value 
$I^{GDH}_{\gamma d \to np}(550\,\mbox{MeV})=7.3\,\mu$b if the nucleon 
anomalous magnetic moment is switched off in the e.m.\ one-body current 
operator (for further details see \cite{Kre96}).

\subsection{Coherent pion production}
\label{sec3b}

The theoretical model used to calculate the contribution of the
coherent pion production channel is described in detail in Refs.\ 
\cite{WiA95} and \cite{WiA96}. The reaction is clearly dominated by the
magnetic dipole excitation of the $\Delta$ resonance from which one expects 
a strong positive $I^{GDH}_{\gamma d \to d\pi^0}$ contribution. The reason 
for this is that the 
$\Delta$-excitation is favoured if photon and nucleon spins are parallel 
compared to the antiparallel situation. The model takes into
account pion rescattering by solving a system of coupled equations for
the N$\Delta$, NN$\pi$ and NN channels.  The most important
rescattering mechanism is due the successive excitation and decay of
the $\Delta$ resonance.  The inclusion of the rescattering effects is 
important and leads in general to a significant reduction of the cross 
section in reasonable agreement with the differential cross 
section data available in the $\Delta$ region. 

\begin{figure}
\centerline{\psfig{figure=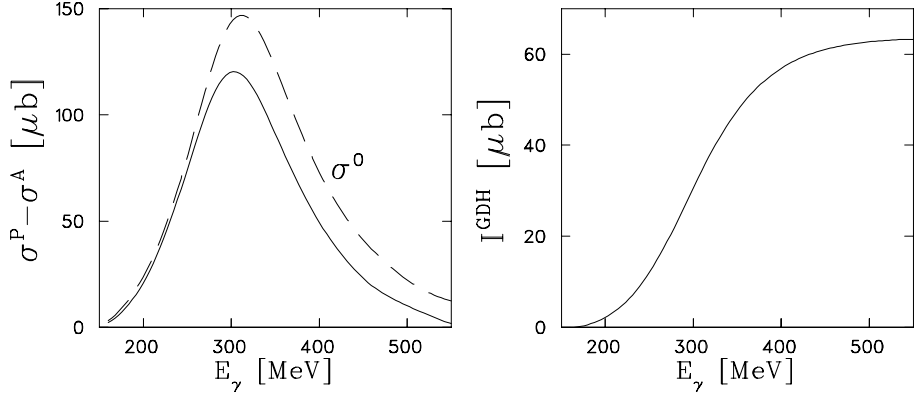,width=12cm,angle=0}}
\vspace*{.5cm}
\caption{
Contribution of the coherent $\pi^0$ production to the GDH sum rule for the 
deuteron. Left panel: difference of the cross sections (full curve), the 
dashed curve shows the unpolarized cross section; right panel: 
$I^{GDH}_{\gamma d \to d\pi^0}$ as function of the upper integration energy.}
\label{fig2}
\end{figure}

Fig.\ \ref{fig2} shows the result of our calculation. One sees the strong 
positive contribution from the $\Delta$-excitation giving a value 
$I^{GDH}_{\gamma d \to d\pi^0}(550\,\mbox{MeV})=63\,\mu$b. The comparison 
with the unpolarized cross section, also plotted in Fig.\ \ref{fig2}, 
demonstrates the dominance of $\sigma^P$ over $\sigma^A$. 
Furthermore, one notes quite satisfactory convergence at the highest 
energy considered here.

\subsection{Incoherent pion production}
\label{sec3c}
The calculation of the $\gamma d \rightarrow \pi NN$ contributions to
the GDH integral is based on the spectator nucleon approach discussed 
in \cite{ScA96}. In this framework, the reaction proceeds 
via the pion production on one nucleon while the other nucleon acts 
merely as a spectator. Thus, the $\gamma d \rightarrow \pi NN$
operator is given as the sum of the elementary $\gamma N \rightarrow \pi N$ 
operators of the two nucleons. For this elementary operator,  
we have taken the standard pseudovector 
Born terms and the contribution of the $\Delta$ resonance, and a satisfactory 
description of pion photoproduction on the nucleon is achieved in the 
$\Delta$-resonance region \cite{ScA96}. Although the spectator model 
does not include any final state interaction except for the resonant 
$M_{1+}^{3/2}$ multipole, it gives quite a good description of available 
data on the total cross section demonstrating the dominance of the 
quasifree production process, for which the spectator model should work 
quite well. 

\begin{figure}
\centerline{\psfig{figure=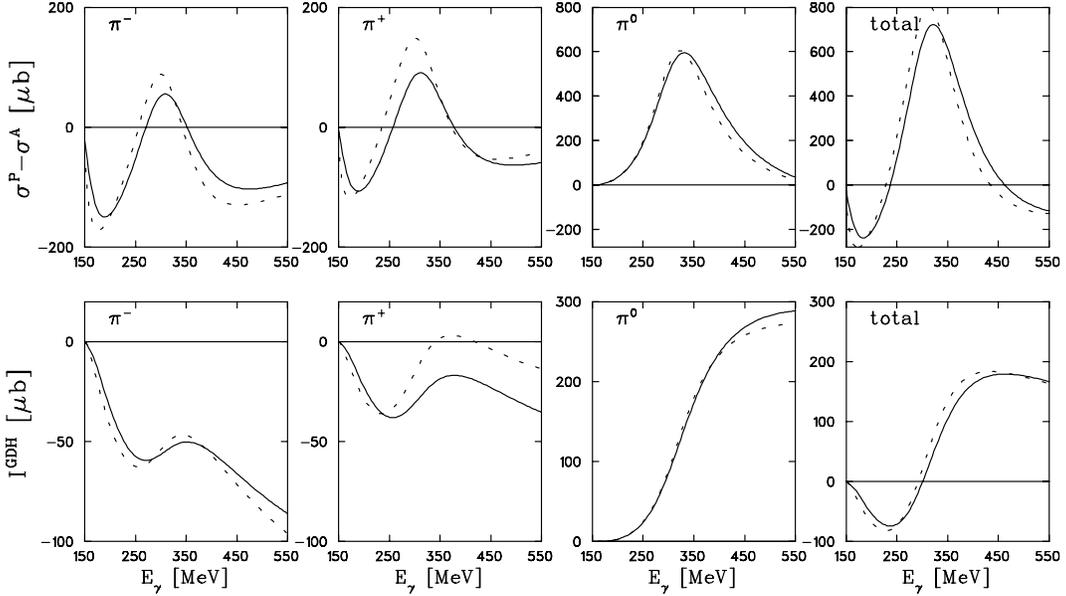,width=14cm,angle=90}}
\vspace*{.5cm}
\caption{
Contribution of the incoherent $\pi$ production to the GDH sum rule for the 
deuteron and the nucleon. Upper part: difference of the cross sections; 
lower part: $I^{GDH}_{\gamma d \to NN\pi}$ as function of the upper integration energy. Full 
curves for the deuteron, dotted curves for the nucleon. In the case of $\pi^0$ 
production, the dotted curve shows the summed proton and neutron 
contributions.}
\label{fig3}
\end{figure}

The results are collected in Fig.\ \ref{fig3}. The upper part shows the 
individual contributions from the different charge states of the pion 
and their total sum to 
the cross section difference for pion photoproduction on both the deuteron 
and for comparison on the nucleon. One notes qualitatively a similar 
behaviour although the maxima and minima are smaller and also slightly 
shifted towards higher energies for the deuteron. In the lower part of 
Fig.\ \ref{fig3} the corresponding GDH integrals are shown. A large positive 
contribution comes from $\pi^0$-production whereas the charged pions give a 
negative but - in absolute size - smaller contribution to the GDH value. Up to 
an energy of 550 MeV one finds for the total contribution of the incoherent 
pion production channels a value 
$I^{GDH}_{\gamma d \to NN\pi}(550\,\mbox{MeV})=167\,\mu$b which is remarkably 
close to the sum of the neutron and proton values for the given elementary 
model $I^{GDH}_n(550\,\mbox{MeV})+I^{GDH}_p(550\,\mbox{MeV})=163\,\mu$b. 
It underlines again that the total cross 
section is dominated by the quasifree process. However, as is 
evident from Fig.\ \ref{fig3}, convergence is certainly not reached at this 
energy. Furthermore, the elementary pion production operator had been 
constructed primarily to give a realistic description of the $\Delta$ 
reonance region. In fact, it underestimates the GDH inegral up to 550 MeV 
by about a factor two compared to a corresponding evaluation based on a 
multipole analysis of experimental pion photoproduction data, as is 
discussed in the next section. For this reason we cannot expect that 
this model gives also a good description of experimental data above 400 MeV. 
But the important result is, that the total GDH contribution from the 
incoherent process is very close to the sum of the free proton and neutron GDH 
integrals which will remain valid for an improved elementary production 
operator. 

\section{Summary and conclusions}
\label{sec4}
The contributions from all three channels and their sum are listed in Tab.\ 
\ref{tab1}. A very interesting and important result is the large negative 
contribution from the photodisintegration channel near and not too far 
above the break-up threshold with surprisingly large relativistic effects 
below 100 MeV. Hopefully, this low energy feature of 
the GDH sum rule could be checked experimentally in the near future. 

For the total GDH value from explicit integration up to 550 MeV, we find a 
negative value $I^{GDH}_d(550\,\mbox{MeV})=-183\,\mu$b. However, as we have 
mentioned above, some uncertainty lies in the contribution of the incoherent
pion production channel because of shortcomings of the model of the
elementary production amplitude above the $\Delta$ resonance. If we use 
instead of the model value $I^{GDH}_{\gamma d \to NN\pi}(550\,\mbox{MeV})=
167\,\mu$b (cf.\ previous section) 
the sum of the GDH values of neutron and proton by integrating the 
cross section difference obtained from a multipole analysis of experimental 
data (fit SM95 from \cite{VPI}), giving $I^{GDH}_n(550\,\mbox{MeV})
+I^{GDH}_p(550\,\mbox{MeV})=331\,\mu$b, we find for the deuteron 
$I^{GDH}_d(550\,\mbox{MeV})=-19\,\mu$b, which we consider a more realistic 
estimate. Since this value is still negative, a positive contribution of 
about the same size should come from contributions at higher energies in 
order to fulfil the small GDH sum rule for the deuteron, provided that the 
sum rule is valid. These contributions should come from the incoherent 
single pion production above 550 MeV because here convergence had not been 
reached in contrast to the other two channels, photodisintegration and 
coherent pion production, and in addition, from multipion production. 

It remains as a task for future research to improve the elementary pion 
photoproduction operator above the $\Delta$ resonance. But for this 
also precise data on $\sigma^P-\sigma^A$ from a direct measurement is 
urgently needed. 
Furthermore, for the reaction on the deuteron, the influence of final state 
interaction has to be investigated, too. Because the large cancellation 
between the various contributions requires quite a high degree of precision 
for the theoretical description. For this reason, also at least two-pion 
production contributions have to be considered in order to obtain more 
reliable predictions at higher energies. 



\begin{table}
\caption{Various contributions of the photodisintegration channel to the 
GDH integral for the deuteron integrated up to 550 MeV in $\mu$b.}
 \begin{tabular}{cccc}
  N & N+MEC & N+MEC+IC & N+MEC+IC+RC\\ 
   \hline
 $-619$ & $-689$ & $-665$ & $-413$ \\
 \end{tabular}
\label{tabdis}
\end{table}

\begin{table}
\caption{Contributions of the different absorption channels to the 
GDH integral for the deuteron integrated up to 550 MeV in $\mu$b.}
 \begin{tabular}{cccccc}
   $\gamma d \to np$ & $\gamma  d \to d \pi^0$ & $\gamma d \to np\pi^0$ 
  &$\gamma d \to nn\pi^+$   & $\gamma d \to pp\pi^-$   &total\\
\hline
  $-413$ &   63 &  288 &  $-35$ &  $-86$ & $-183$ \\
 \end{tabular}
\label{tab1}
\end{table}

\newpage

\end{document}